\documentclass[english,twocolumn, 10pt]{IEEEtran}
\usepackage[T1]{fontenc}
\usepackage[latin9]{inputenc}
\usepackage{float}
\usepackage{bm}
\usepackage{amsmath}
\usepackage{amsthm}
\usepackage{amssymb}
\usepackage{graphicx}

\makeatletter

\floatstyle{ruled}
\newfloat{algorithm}{tbp}{loa}
\providecommand{\algorithmname}{Algorithm}
\floatname{algorithm}{\protect\algorithmname}

\theoremstyle{plain}
\newtheorem{thm}{\protect\theoremname}
\theoremstyle{plain}
\newtheorem{lem}[thm]{\protect\lemmaname}

\usepackage{bm}
\usepackage{subfigure}
\usepackage{cite}
\usepackage[english]{babel}
\usepackage[T1]{fontenc}
\usepackage{algorithm}
\usepackage{algorithmic}
\makeatletter
\def\blfootnote{\xdef\@thefnmark{}\@footnotetext}
\makeatother

\makeatother

\usepackage{babel}
\providecommand{\lemmaname}{Lemma}
\providecommand{\theoremname}{Theorem}

\begin{document}

\title{Age-Optimal Trajectory Planning for UAV-Assisted Data Collection\thanks{This work was supported in part by the National Natural Science Foundation
of China (NSFC) under Grant No. 61601255, and the Scientific Research
Foundation of Ningbo University under Grant No. 010-421703900. }}

\author{Juan Liu$^{\dagger}$, Xijun Wang$^{\dagger\dagger}$, \IEEEmembership{Member, IEEE}
and Bo Bai$^{*}$, \IEEEmembership{Senior Member, IEEE}, Huaiyu Dai$^{\diamondsuit}$,
\IEEEmembership{Fellow, IEEE},\\ $^{\dagger}$College of Electrical
Engineering and Computer Science, Ningbo University, Zhejiang 315211,
China\\$^{\dagger\dagger}$State Key Laboratory of Integrated Service
Networks, Institute of Information Science\\ Xidian University, Xi\textquoteright an,
Shaanxi, 710071, China\\ $^{*}$Future Network Theory Lab, 2012 Labs,
Huawei Technologies Co., Ltd., Shatin, N. T., Hong Kong\\$^{\diamondsuit}$Department
of Electrical and Computer Engineering, North Carolina State University,
USA\\Email: eeliujuan@gmail.com,$\,$xjwang22@gmail.com,$\,$baibo8@huawei.com,$\,$hdai@ncsu.edu}
\maketitle
\begin{abstract}
Unmanned aerial vehicle (UAV)-aided data collection is a new and promising
application in many practical scenarios. In this work, we study the
age-optimal trajectory planning problem in UAV-enabled wireless sensor
networks, where a UAV is dispatched to collect data from the ground
sensor nodes (SNs). The age of information (AoI) collected from each
SN is characterized by the data uploading time and the time elapsed
since the UAV leaves this SN. We attempt to design two age-optimal
trajectories, referred to as the Max-AoI-optimal and Ave-AoI-optimal
trajectories, respectively. The Max-AoI-optimal trajectory planning
is to minimize the age of the `oldest' sensed information among the
SNs.  The Ave-AoI-optimal trajectory planning is to minimize the average
AoI of all the SNs. Then, we show that each age-optimal flight trajectory
corresponds to a shortest Hamiltonian path in the wireless sensor
network where the distance between any two SNs represents the amount
of inter-visit time. The dynamic programming (DP) method and genetic
algorithm (GA) are adopted to find the two different age-optimal trajectories.
Simulation results validate the effectiveness of the proposed methods,
and show how the UAV's trajectory is affected by the two  AoI metrics.
\end{abstract}

\section{Introduction}

Unmanned aerial vehicles (UAVs) have attracted a lot of attentions
in both academia and industry \cite{2016_NHMotlagh_IOT}. Thanks to
its fully controllable mobility, the UAV can be employed to collect
the sensed data from the ground sensor nodes (SNs) as a mobile data
collector. The UAV can move very close to each SN and communicate
with it via low-altitude line-of-sight (LoS) communication links \cite{2017_RuiZhang_arkiv}.
Hence, the UAV-assisted data collection can save the transmission
energy of each SN, and prolong the lifetime of wireless sensor networks. 

The efficiency of data collection and design of UAV\textquoteright s
trajectory have been studied recently for UAV-assisted wireless sensor
networks \cite{2014_Abdulla_Infocom,2016_SSay_UAV,2017_Rzhang_UAV_data_collection}.
In \cite{2014_Abdulla_Infocom}, considering the fairness between
the cluster heads that communicate with the UAV directly, Abdulla
et al. formulated the energy efficiency maximization problem in UAV-aided
data collection as a potential game. The impacts of the UAV's trajectory
on the adaptive modulation scheme were also discussed. Say et al.
proposed in \cite{2016_SSay_UAV} a UAV-assisted data gathering framework
for wireless sensor networks, where the UAV is used as a relay to
collect the sensed data from the SNs. Inside the UAV's coverage area,
the sensors are divided into different frames according to their locations
and assigned with different transmission priorities to reduce the
packet loss. The authors of \cite{2017_Rzhang_UAV_data_collection}
studied the joint optimization of the SNs' wake-up schedule and UAV's
trajectory in UAV-enabled wireless sensor networks for reliable and
energy-efficient data collection. The successive convex optimization
method was applied to find a sub-optimal UAV's trajectory. It is shown
in these works that the UAV's trajectory greatly affects the efficiency
of data collection. 

On the other hand, the freshness of the sensed information is very
crucial in delay-sensitive applications. The freshness is a new and
important metric, referred to as the age of information (AoI) or status
age, which is defined as the amount of time elapsed since the instant
at which the freshest delivered update takes place \cite{2012_SKaul_AoI_Infocom}.
Most existing works focused on analyzing or optimizing the AoI-based
schedule or transmission \cite{2012_SKaul_AoI_Infocom,2013_CKam_ISIT,2017_YSun_aoi_IT}.
In \cite{2012_SKaul_AoI_Infocom}, the age-optimal throughput region
was derived for the first-come-first-served M/M/1 queueing system.
The status age was derived for status updates randomly generated and
transmitted in a cloud-based network in \cite{2013_CKam_ISIT}. Sun
et al. studied the optimal control policy of information updates in
a communication system and proposed the zero-wait policy to keep the
data fresh \cite{2017_YSun_aoi_IT}. Efficient algorithms were developed
to find the optimal update policy in the constrained semi-Markov decision
framework. 

Motivated by the above works, we are interested in investigating the
impacts of AoI metrics on the design of UAV-assisted data collection.
In this work, we study the age-optimal trajectory planning problem
in UAV-enabled wireless sensor networks. The UAV takes off from the
data center, collects the sensed data from all the SNs sequentially
and delivers them to the data center for information processing when
it flies back to it. The AoI of each SN is equal to the amount of
time elapsed from the instant at which the information is sensed to
the instant at which the information is delivered to the data center.
It is easy to see that the UAV's trajectory has a great impact on
the AoI of each SN. Particularly, we define two different AoI metrics,
i.e., the maximum AoI and average AoI of SNs. Accordingly, the UAV's
optimal trajectories are referred to as the Max-AoI-optimal and Ave-AoI-optimal
trajectories, respectively. Through theoretical analysis, we show
that the Max-AoI-optimal trajectory is exactly a shortest Hamiltonian
path while the Ave-AoI-optimal one is a stage-weighted shortest Hamiltonian
path in the wireless sensor network. Then, we adopt the dynamic programming
(DP) approach to find the two age-optimal trajectories recursively.
The computational complexity of DP is rather high as the network size
increases. For large-scale wireless sensor networks, we develop a
genetic algorithm (GA) to intelligently search the near optimal trajectories.
Simulation results show the importance of the two AoI metrics in the
design of the UAV's trajectory. 

\section{System Model\label{sec:System-Model}}

\begin{figure}[t]
\centering
\renewcommand{\figurename}{Fig.}

\includegraphics[width=0.43\textwidth]{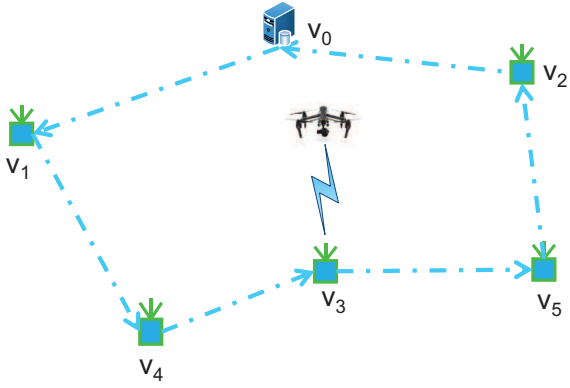}\caption{An illustrative model of UAV-enabled data collection: The UAV flies
following a trajectory $v_{0}\rightarrow v_{1}\rightarrow v_{4}\rightarrow v_{3}\rightarrow v_{5}\rightarrow v_{2}\rightarrow v_{0}$,
collects the latest sensing information from each SN $v_{i}$ ($i=1,\cdots,5$),
and flies back to the data center $v_{0}$. }
\label{fig:system_model}
\end{figure}
As shown in Fig.$\,$\ref{fig:system_model}, we consider a UAV-enabled
wireless sensor network in which a UAV is dispatched to collect data
from $M$ SNs, denoted by $\mathcal{V}=\left\{ v_{1},\cdots,v_{M}\right\} $,
and then flies back to the data center $v_{0}$ for data analysis.
The network can be described by a complete graph $G=(\mathcal{V}^{+},\,\mathcal{E})$,
where $\mathcal{V}^{+}=\left\{ v_{0}\right\} \bigcup\mathcal{V}$
denotes the set of all the nodes including the data center, and $\mathcal{E}=\{(v_{i},v_{j})\,|v_{i},v_{j}\in\mathcal{V},\,i\neq j\}$
denotes the set of all the edges connecting any two nodes. For ease
of exposition, we denote by $\mathcal{N}_{m}=\{v_{i}|(v_{i},v_{m})\in\mathcal{E},v_{i}\in\mathcal{V}^{+}\}$
the set of the neighboring nodes of node $v_{m}$. Each edge $(v_{i},v_{j})$
has an associated non-negative weight $d_{i,j}$, indicating the distance
between nodes $v_{i}$ and $v_{j}$, i.e., $d_{i,j}=\Vert\bm{s}_{i}-\bm{s}_{j}\Vert$,
where $\bm{s}_{i}=[x_{i},y_{i}]\in\mathbb{R}^{2}$ is the location
of node $v_{i}$.

To collect the information, the UAV takes off from the data center
$v_{0}$, flies to collect the information from all the $M$ nodes
following a prescheduled trajectory, and lands at the data center
after completing the job. The UAV flies at a fixed altitude $h$ and
maintains a constant flight velocity, denoted by $V$. Except the
start/end point $v_{0}$, the trajectory is supposed to contain the
sequence of the non-repeating nodes, like $v_{0}\rightarrow v_{(1)}\rightarrow v_{(2)}\rightarrow\cdots\rightarrow v_{(M)}\rightarrow v_{(M+1)}=v_{0}$,
 where $v_{(i)}\in\mathcal{V}$ denotes the $i$-th node in the trajectory.
Here, the trajectory vector $\bm{u}=[v_{(1)},v_{(2)}\cdots,v_{(M)}]$
specifies one permutation of the node set $\mathcal{V}$. 

When the UAV flies to and hovers above node $v_{(i)}$, it establishes
a line-of-sight communication link immediately with this node. Node
$v_{(i)}$ samples its sensing information, packs it in a data packet
of length $L_{(i)}^{p}$ with timestamp $T_{i}$, and transmits the
tagged packet to the UAV. Without loss of generality, we assume that
the UAV takes off at time $T_{0}=0$. The channel power gain of the
LOS link from the node to the UAV can be modeled as $g=\beta h^{-2}$,
where $\beta$ denotes the channel gain at the reference distance
\cite{2017_RuiZhang_arkiv}. When the node transmits at a constant
power $P_{(i)}$, its uploading data rate over the LOS link can be
expressed as
\begin{equation}
R_{(i)}=B\log_{2}\left(1+\frac{g\cdot P_{(i)}}{\sigma^{2}}\right)=B\log_{2}\left(1+\frac{\beta}{h^{2}\sigma^{2}}P_{(i)}\right),
\end{equation}
where $B$ denotes the system bandwidth, and $\sigma^{2}$ denotes
the noise power at the UAV receiver. Assuming that the sensing time
is very small and negligible, the data collection time at node $v_{(i)}$
can be evaluated as $t_{(i)}^{tx}=L_{(i)}^{p}/R_{(i)}$. The set of
SNs that have been visited by the UAV till now is denoted by $\mathcal{C}_{i}=\{v_{(1)},\cdots,v_{(i)}\}$.
Then, the UAV flies directly to the next node $v_{(i+1)}$ in the
trajectory and continues the data collecting job. After visiting all
the nodes, the UAV returns back to the data center at time $T_{(M+1)}$
for data analysis.

We use $X_{i}(t)$ to track the age of the information collected from
the $i$-th node $v_{(i)}$ in the flight trajectory at time $t$.
When $t<T_{i}$, $X_{i}(t)=0$, since node $v_{(i)}$ has not been
visited at this time and its information has not been sampled \cite{2012_SKaul_AoI_Infocom}.
Otherwise, $X_{i}(t)=t-T_{i}$. Hence, the information age of node
$v_{(i)}$ is given by 
\begin{equation}
X_{i}(t)=\left(t-T_{i}\right)^{+},\quad i=1,2,\cdots,M,
\end{equation}
where $(x)^{+}=\max\{0,x\}.$ At instant $t=T_{i+1}$, the AoI collected
from node $v_{(i)}$ is equal to 
\begin{equation}
X_{i}(T_{i+1})\triangleq\eta_{(i),(i+1)}=t_{(i)}^{tx}+t_{(i),(i+1)}^{flight},\label{eq:eta_ij}
\end{equation}
where $t_{(i),(i+1)}^{flight}=V^{-1}d_{(i),(i+1)}$ denotes the flight
time of the UAV from node $v_{(i)}$ to node $v_{(i+1)}$. Here, $\eta_{(i),(i+1)}$
indicates the amount of time elapsed between two data collection actions
at node $v_{(i)}$ and node $v_{(i+1)}$. When getting back to the
data center at time $T_{(M+1)}$, the UAV has visited all the SNs
and collected $M$ data packets including sensing information with
different ages, given by
\begin{equation}
X_{i}(T_{(M+1)})=T_{(M+1)}-T_{i}=\sum_{k=i}^{M}\eta_{(k),(k+1)},\label{eq:age_node_i}
\end{equation}
which is totally determined by the trajectory $\bm{u}$ and hence
can be expressed as a function of the trajectory $\bm{u}$, i.e.,
$X_{i}(T_{(M+1)})=X_{i}(\bm{u})$. Similarly, the average age of the
sensing information can be computed as

\begin{equation}
\overline{X}(T_{(M+1)})=\overline{X}(\bm{u})=\frac{1}{M}\sum_{i=1}^{M}X_{i}(\bm{u}).\label{eq:ave_age_nodes}
\end{equation}

In this work, we attempt to design two age-optimal flight trajectories
for UAV-assisted data collection. One is to minimize the age of the
`oldest' sensing information among the SNs. The other is to minimize
the average age of the sensing information. In particular, we formulate
two combinatorial optimization problems as follows:

\begin{equation}
\mathcal{P}_{1}:\quad\min_{\bm{u}}\,\max_{i\in\{1,\cdots,M\}}\left\{ X_{i}(\bm{u})\right\} \label{eq:maximum_AOI}
\end{equation}
and 
\begin{equation}
\mathcal{P}_{2}:\quad\min_{\bm{u}}\,\overline{X}(\bm{u}).\label{eq:average_AOI}
\end{equation}
The optimal solutions to Problems $\mathcal{P}_{1}$ and $\mathcal{P}_{2}$
are denoted by $\bm{u}_{max}^{*}$ and $\bm{u}_{ave}^{*}$, referred
to as the optimal maximum-age-of-information (Max-AoI-optimal) trajectory
and the optimal average-age-of-information (Ave-AoI-optimal) trajectory,
as discussed below. 

\section{AoI-Optimal Trajectory Planning}

By analyzing the properties of Problems $\mathcal{P}_{1}$ and $\mathcal{P}_{2}$,
we show that the Max-AoI-optimal and Ave-AoI-optimal flight trajectories
are actually two shortest Hamiltonian paths. 

\subsection{Max-AoI-optimal Trajectory}

Among all the SNs, the first node $v_{(1)}$ in the trajectory $\bm{u}$
always experiences the largest AoI, as presented in the following
lemma. 
\begin{lem}
\label{lem:flight_max_aoi}For any flight trajectory $\bm{u}$, we
have
\begin{equation}
X_{1}(\bm{u})>X_{2}(\bm{u})>\cdots>X_{M}(\bm{u}).\label{ieq:age_node}
\end{equation}
\end{lem}
From Lemma \ref{lem:flight_max_aoi}, Problem $\mathcal{P}_{1}$ is
equivalent to 
\begin{equation}
\mathcal{P}_{1}^{'}:\quad\min_{\bm{u}}\quad X_{1}(\bm{u}).\label{eq:maximum_AOI_X1}
\end{equation}
Accordingly, the optimal value of Problem $\mathcal{P}_{1}^{'}$ is
denoted by $X_{1}^{*}=X_{1}(\bm{u}_{max}^{*})$. 
\begin{thm}
The Max-AoI-optimal trajectory is a shortest Hamiltonian path that
where the distance between any two nodes $v_{i}$ and $v_{j}$ is
equal to $\eta_{i,j}$.
\end{thm}
\begin{IEEEproof}
Notice that the length of the flight trajectory is equal to

\begin{equation}
X_{1}(\bm{u})=\sum_{k=1}^{M}\eta_{(i),(i+1)}.\label{eq:age_X1}
\end{equation}
Thus, solving Problem $\mathcal{P}_{1}^{'}$ is equivalent to finding
 a shortest Hamiltonian path $\bm{u}_{max}^{*}$ that starts from
node $v_{(1)}$ and visits all the other nodes exactly once before
going back to the data center $v_{0}$. The AoI  of node $v_{(1)}$
is equal to the length of the Hamiltonian path $X_{1}(\bm{u})$, which
is to be minimized. 
\end{IEEEproof}
In graph theory, a Hamiltonian path is a path in an undirected or
directed graph that visits each vertex exactly once \cite{2016_RDiestel_graphtheory}.
There exists a Hamiltonian path in the complete graph $G$. Hence,
to find the Max-AoI-optimal flight trajectory, we shall find a shortest
Hamiltonian path that visits all the SNs $\{v_{i}\}$ $(i=1,\cdots,M)$
exactly once and goes back to the data center $v_{0}$. In this scenario,
the time parameters $\{\eta_{i,j}\}$ $(\forall i,j\in\mathcal{V}^{+})$
are treated as the distances between any two nodes. 

\subsection{Ave-AoI-optimal Trajectory Planning}

Similarly, we analyze the average age of the sensing information $\overline{X}(\bm{u})$
and discuss the Ave-AoI-optimal trajectory planning problem $\mathcal{P}_{2}$. 
\begin{lem}
The average AoI given by (\ref{eq:ave_age_nodes}) can be re-expressed
as

\begin{equation}
\overline{X}(\bm{u})=\sum_{i=1}^{M}\frac{i}{M}\cdot\eta_{(i),(i+1)}.\label{eq:ave_age_nodes_v1}
\end{equation}
\end{lem}
For ease of exposition, we define the weighted AoI collected from
node $v_{(i)}$ to node $v_{(M)}$ along the trajectory $\bm{u}$
as
\begin{equation}
\overline{X}_{i}(\bm{u})=\sum_{k=i}^{M}\frac{k}{M}\cdot\eta_{(k),(k+1)},\label{eq:ave_age_nodes_vi}
\end{equation}
which satisfies 
\begin{equation}
\overline{X}_{i}(\bm{u})>\overline{X}_{j}(\bm{u})\quad\forall i<j.\label{eq:average_aoi_inequality}
\end{equation}
 Thus, the average information age is obtained as $\overline{X}(\bm{u})=\overline{X}_{1}(\bm{u})$.
Thus, Problem $\mathcal{P}_{2}$ is equivalent to 
\begin{equation}
\mathcal{P}_{2}^{'}:\quad\min_{\bm{u}}\quad\overline{X}_{1}(\bm{u}).\label{eq:min_average_AOI}
\end{equation}
The Ave-AoI-optimal trajectory is the optimal solution of $\mathcal{P}_{2}^{'}$
and the optimal value is denoted by $\overline{X}_{1}^{*}=\overline{X}_{1}(\bm{u}_{ave}^{*})$
accordingly.
\begin{thm}
The Ave-AoI-optimal trajectory is a stage-weighted shortest Hamiltonian
path where the stage-weighted distance between node $v_{(k)}$ and
$v_{(k+1)}$ is equal to $\frac{k}{M}\cdot\eta_{(k),(k+1)}$. 
\end{thm}
\begin{IEEEproof}
For any trajectory $\bm{u}$, the average AoI can be viewed as the
stage-weighted length of the trajectory, i.e.,
\begin{equation}
\overline{X}_{1}(\bm{u})=\sum_{k=1}^{M}\frac{k}{M}\cdot\eta_{(k),(k+1)},\label{eq:average_age_X1}
\end{equation}
where $\frac{k}{M}$ is the weight. Specifically, the distance between
nodes $v_{(k)}$ and $v_{(k+1)}$ is equal to $\frac{k}{M}\cdot\eta_{(k),(k+1)}$.
Thus, solving Problem $\mathcal{P}_{2}^{'}$ is equivalent to finding
a stage-weighted shortest Hamiltonian path $\bm{u}_{ave}^{*}$ that
starts from node $v_{(1)}$ and visits all the other nodes exactly
once before going back to the data center $v_{0}$. From (\ref{eq:average_age_X1}),
the minimum average AoI is equal to the stage-weighted length of the
shortest Hamiltonian path $\bm{u}_{ave}^{*}$. 
\end{IEEEproof}
In this theorem, we show that the Ave-AoI-optimal trajectory can also
be regarded as a shortest Hamiltonian path in the wireless sensor
network. However, the distance between two consecutive nodes $v_{(k)}$
and $v_{(k+1)}$ in the trajectory is equal to the product of the
factor $\frac{k}{M}$ and the time parameter $\eta_{(k),(k+1)}$.
In the sequel, we discuss the algorithm design problem for AoI-Optimal
trajectory planning. 

\section{Algorithm Design for AoI-Optimal Trajectory Planning}

In this part, we discuss how to find the Max-AoI-optimal and Ave-AoI-optimal
trajectories using methods of DP and genetic algorithm (GA), respectively.
\begin{algorithm}[t]
\caption{DP-based age-optimal trajectory planning \label{alg:DP_AoI_optimal}}
\begin{algorithmic}[1]

\STATE $\textbf{Input}$: the network topology $G=(\mathcal{V}^{+},\,\mathcal{E})$,
and the system parameters $(V,h,\beta,B,P_{m},\sigma^{2})$ for all
$v_{m}\in\mathcal{V}$.

\STATE Calculate the elapsed time of data collection between any
two nodes $\eta_{i,j}$ (c.f. (\ref{eq:eta_ij})) for all $i\neq j$,
$v_{i},v_{j}\in\mathcal{V}^{+}$;

\FOR{$i=1:M$}

\FOR{$S\subseteq\mathcal{V}-\{v_{i}\}$}

\STATE Calculate the minimum path cost $f(i,S)$ (or $g(i,S)$) according
to (\ref{eq:dp_max_aoi}) (or (\ref{eq:dp_ave_aoi}));

\ENDFOR

\ENDFOR

\STATE Calculate the minimum AoI as $X_{1}^{*}$ (or $\overline{X}_{1}^{*}$)
by (\ref{eq:dp_aoi_node1}) (or (\ref{eq:dp_aoi_ave_node1}));

\STATE Find the optimal node $v_{(1)}^{*}=\arg\min_{v_{i}\in\mathcal{V}}f(i,\mathcal{V}-\{v_{i}\})$
(or $v_{(1)}^{*}=\arg\min_{v_{i}\in\mathcal{V}}g(i,\mathcal{V}-\{v_{i}\})$)
that is firstly visited;

\STATE Trace back to find the optimal trajectory $\bm{u}_{max}^{*}$
(or $\bm{u}_{ave}^{*}$) starting with node $v_{(1)}^{*}$ and ending
with node $v_{0}$; 

\STATE $\textbf{Output}$: the optimal trajectory $\bm{u}_{max}^{*}$
(or $\bm{u}_{ave}^{*}$).

\end{algorithmic}
\end{algorithm}

\subsection{DP-based AoI-Optimal Trajectory Planning}

\textbf{i) The Max-AoI-optimal Case: }We first apply the DP method
to find the Max-AoI-optimal flight trajectory in the wireless sensor
network. Let $f(i,S)$ $(v_{i}\notin S\subset\mathcal{V})$ denote
the minimum cost of the path starting from node $v_{i}$, passing
all the vertices in the set $S$ exactly once and going back to the
data center $v_{0}$. The minimum path cost $f(i,S)$ can be expressed
in a recursive form: 
\begin{equation}
f(i,S)=\begin{cases}
\eta_{i,0}, & S=\emptyset,\\
\min\limits _{v_{k}\in S}\left\{ \eta_{i,k}+f(k,S-\{k\})\right\} , & S\neq\emptyset.
\end{cases}\label{eq:dp_max_aoi}
\end{equation}
The Max-AoI-optimal trajectory is exactly the shortest Hamiltonian
path that achieves the minimum cost
\begin{equation}
X_{1}^{*}=\min_{v_{i}\in\mathcal{V}}f(i,\mathcal{V}-\{v_{i}\}),\label{eq:dp_aoi_node1}
\end{equation}
where the cost function $f(i,\mathcal{V}-\{v_{i}\})$ is calculated
by (\ref{eq:dp_max_aoi}). 

\textbf{ii) The Ave-AoI-optimal Case: }Similarly, we show that the
Ave-AoI-optimal trajectory can also be found using the DP method.
Let $g(i,S)$ $(v_{i}\notin S)$ denote the minimum weighted cost
of the path starting from node $v_{i}$, passing all the vertices
in the set $S$ exactly once and returning back to the data center
$v_{0}$. To find the stage-weighted shortest Hamiltonian path, we
express the minimum weighted path cost $g(i,S)$ as: 
\begin{equation}
g(i,S)=\begin{cases}
\eta_{i,0}, & S=\emptyset,\\
\min\limits _{v_{k}\in S}\left\{ (1-\frac{|S|}{M})\eta_{i,k}+g(k,S-\{k\})\right\} , & S\neq\emptyset.
\end{cases}\label{eq:dp_ave_aoi}
\end{equation}
where $|S|$ denotes the cardinality of the set $S$, i.e., the number
of elements in $S$. Here, $|S|$ is used to indicate how many nodes
remain unvisited. Accordingly, $1-\frac{|S|}{M}$ can be used to represent
the stage weight. Therefore, the Ave-AoI-optimal trajectory is also
the `shortest' Hamiltonian path that obtains the minimum path cost
as
\begin{equation}
\overline{X}_{1}^{*}=\min_{v_{i}\in\mathcal{V}}g(i,\mathcal{V}-\{v_{i}\}),\label{eq:dp_aoi_ave_node1}
\end{equation}
where the cost function $g(i,\mathcal{V}-\{v_{i}\})$ is calculated
by (\ref{eq:dp_ave_aoi}) iteratively. 

\textbf{iii) Algorithm Design: }According to the above discussions,
we propose a DP based algorithm, i.e., Algorithm \ref{alg:DP_AoI_optimal},
to find the two AoI-optimal trajectories. In particular, we first
calculate the distance $\eta_{i,j}$ between any two nodes $v_{i}$
and $v_{j}$ by (\ref{eq:eta_ij}). To find the Max-AoI-optimal trajectory,
we calculate the minimum path costs $f(i,S)$ recursively by (\ref{eq:dp_max_aoi})
for each node $v_{i}\in\mathcal{V}$ and all the subsets $S\subseteq\mathcal{V}-\{v_{i}\}$.
The minimum path cost $f(i,S)$ and the node $\tau\in S$ being visited
exactly after node $v_{i}$ are recorded in a table. From (\ref{eq:dp_aoi_node1}),
the minimum AoI as $X_{1}^{*}$ is found by comparing the minimum
path costs $f(i,\mathcal{V}-\{v_{i}\})$ for all $v_{i}\in\mathcal{V}$.
Accordingly, the optimal node firstly visited by the UAV is marked
as $v_{(1)}^{*}$. The shortest Hamiltonian path $\bm{u}_{max}^{*}$
can be found by tracing back the data stored in the table. Using the
DP algorithm, the Ave-AoI-optimal trajectory $\bm{u}_{ave}^{*}$ can
be found in the same way. Notice that the minimum weighted path costs
$g(i,S)$ (c.f. (\ref{eq:dp_ave_aoi})) rather than the path costs
$f(i,S)$ shall be calculated and recorded. 

In this way, we can find the Max-AoI-optimal and Ave-AoI-optimal trajectories
using Algorithm \ref{alg:DP_AoI_optimal}. The computational complexity
of the DP method is about $O(M\cdot2^{M})$, which becomes intolerable
when the number of nodes $M$ is large. To deal with this issue, we
develop a genetic algorithm (GA) to solve the AoI-optimal trajectory
planning problems $\mathcal{P}_{1}^{'}$ and $\mathcal{P}_{2}^{'}$
in the sequel.

\subsection{GA-based AoI-Optimal Trajectory Planning}

\begin{algorithm}[t]
\caption{GA-based age-optimal trajectory planning \label{alg:GA_AoI_optimal}}
\begin{algorithmic}[1]

\STATE $\textbf{Input}$: the network topology $G=(\mathcal{V}^{+},\,\mathcal{E})$,
the data collecting time matrix $\bm{\eta}=[\eta_{i,j}]$, and the
GA related parameters $\{N_{c},N_{g},\alpha,\gamma_{c},\gamma_{m}\}$;

\STATE \textbf{Step 1}: Create an initial population of $N_{c}$
chromosomes, and set the generation index as $n=0$;

\STATE \textbf{Step 2}: Evaluate the normalized fitness of each chromosome
$\bm{u}$ by substituting $l(\bm{u})=X_{1}(\bm{u})$ or $l(\bm{u})=\overline{X}_{1}(\bm{u})$
into (\ref{eq:fitness});

\STATE \textbf{Step 3}: Randomly choose the parent chromosomes via
proportional selection according to the threshold $\gamma_{c}$; Create
offsprings from the selected parent chromosomes using partially mapped
crossover operation; Apply the mutation operations on the current
population of $N_{c}$ chromosomes with probability $\gamma_{m}$; 

\STATE \textbf{Step 4}: Generate a new population of $N_{c}$ chromosomes
by replacing old chromosomes with new ones, and increase the generation
index $n$ by one, i.e., $n=n+1$; The procedure goes to \textbf{Step
2} if $n<N_{g}$; Otherwise, it is terminated.

\STATE \textbf{Step 5}: Calculate the fitness of each chromosome
$\bm{u}\in\mathcal{C}$, and find the optimal trajectory $\bm{u}_{max}^{*}=\arg\max\limits _{\bm{u}\in\mathcal{C}}l(\bm{u})=X_{1}(\bm{u})$
(or $\bm{u}_{ave}^{*}=\arg\max\limits _{\bm{u}\in\mathcal{C}}l(\bm{u})=\overline{X}_{1}(\bm{u})$);

\STATE $\textbf{Output}$: the optimal trajectory $\bm{u}_{max}^{*}$
(or $\bm{u}_{ave}^{*}$).

\end{algorithmic}
\end{algorithm}
In Algorithm \ref{alg:GA_AoI_optimal}, we present the GA-based AoI-optimal
trajectory planning algorithm which consists of the following several
steps.

\textbf{Step 1:} we create an initial population of $N_{c}$ chromosomes,
denoted by a set $\mathcal{C}$. Each chromosome represents a feasible
Hamiltonian path $\bm{u}$ in the wireless sensor network. The generation
index $n$ is set as zero. 

\textbf{Step 2:} we evaluate the normalized fitness of each chromosome
$\bm{u}$ as
\begin{equation}
\phi(\bm{u})=\left(1-\frac{l(\bm{u})-l_{min}(\mathcal{C})}{l_{max}(\mathcal{C})-l_{min}(\mathcal{C})+\epsilon}\right)^{\alpha},\label{eq:fitness}
\end{equation}
where $l(\bm{u})$ denotes the length of path $\bm{u}$ that is equal
to the maximum age $X_{1}(\bm{u})$ or the maximum average age $\overline{X}_{1}(\bm{u})$,
$l_{max}(\mathcal{C})=\max_{\bm{u}\in\mathcal{C}}l(\bm{u})$ and $l_{min}(\mathcal{C})=\min_{\bm{u}\in\mathcal{C}}l(\bm{u})$
are the maximum length and the minimum length of all the paths in
the set $\mathcal{C}$, $\alpha$ is an acceleration factor that is
larger than one, and $\epsilon$ is a very small value used for adjustment. 

\textbf{Step 3:} we apply the selection, crossover and mutation operations
on the population of $N_{c}$ chromosomes. Firstly, we randomly choose
the parent chromosomes via proportional selection for mating. This
means that in accordance with shorter Hamiltonian paths, the chromosomes
with higher fitness have a larger probability of being selected. Specifically,
the chromosome is selected if its fitness value is larger than a threshold
$\gamma_{c}$. The set of the selected chromosomes is denoted by $N_{c}^{'}$.
Secondly, we randomly choose $\lfloor\frac{N_{c}^{'}}{2}\rfloor$
pairs of parents to create offsprings using partially mapped crossover
operation. By marking two random cut points on a pair of parent paths,
the nodes between cut points are exchanged between the two paths to
create two offspring paths. The repetitive nodes between the offspring
paths are then exchanged to produce two offspring Hamiltonian paths.
Thirdly, we apply mutation operation on each chromosome with probability
$\gamma_{m}$. That is, two nodes of a path $\bm{u}$ are randomly
selected and exchanged several time. 

\textbf{Step 4:} we generate a new population of $N_{c}$ chromosomes
by replacing a proportion of old chromosomes with new ones. Thus,
the number of generations $n$ is increased by one. The procedure
is terminated if the number of generations reaches its maximum value
$N_{g}$, i.e., $n=N_{g}$. Otherwise, the process goes to \textbf{Step
2} and repeats. 

\textbf{Step 5:} we calculate the fitness of each chromosome in the
final generation, and set the chromosome with the maximum fitness
as the optimal flight trajectory, i.e., $\bm{u}_{max}^{*}=\arg\max\limits _{\bm{u}\in\mathcal{C}}l(\bm{u})=X_{1}(\bm{u})$
(or $\bm{u}_{ave}^{*}=\arg\max\limits _{\bm{u}\in\mathcal{C}}l(\bm{u})=\overline{X}_{1}(\bm{u})$);

In this way, we develop the GA-based algorithm to find the Max-AoI-optimal
and Ave-AoI-optimal trajectories.  
\begin{figure}[t]
\centering \includegraphics[width=0.43\textwidth]{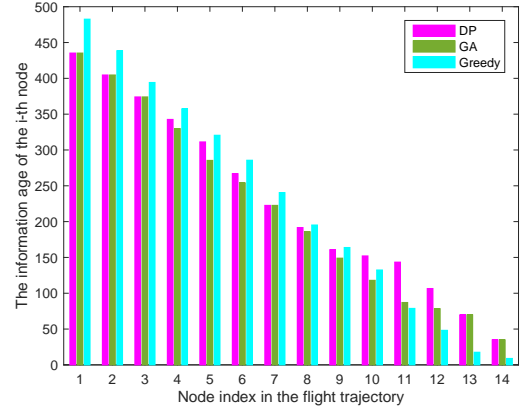}\caption{The minimum AoI $X_{i}^{*}$ in the Max-AoI-optimal trajectory with
$M=14$ SNs. \label{fig:stage_Max_AoI}}
\end{figure}
\begin{figure}[t]
\centering \includegraphics[width=0.43\textwidth]{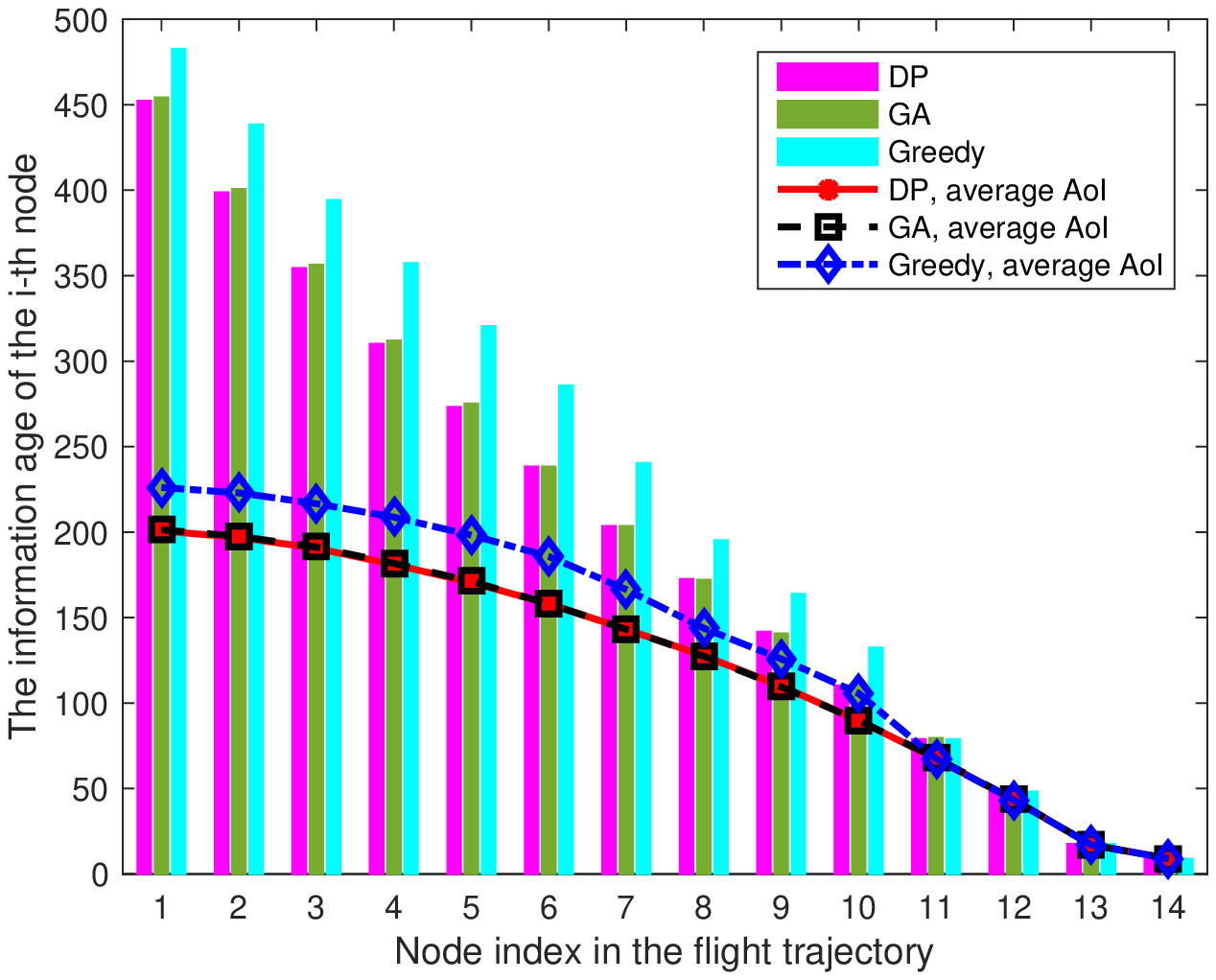}\caption{The minimum average AoI $\overline{X}_{i}^{*}$ in the Ave-AoI-optimal
trajectory with $M=14$ SNs. \label{fig:stage_Ave_AoI}}
\end{figure}
\begin{figure}[t]
\centering \includegraphics[width=0.42\textwidth]{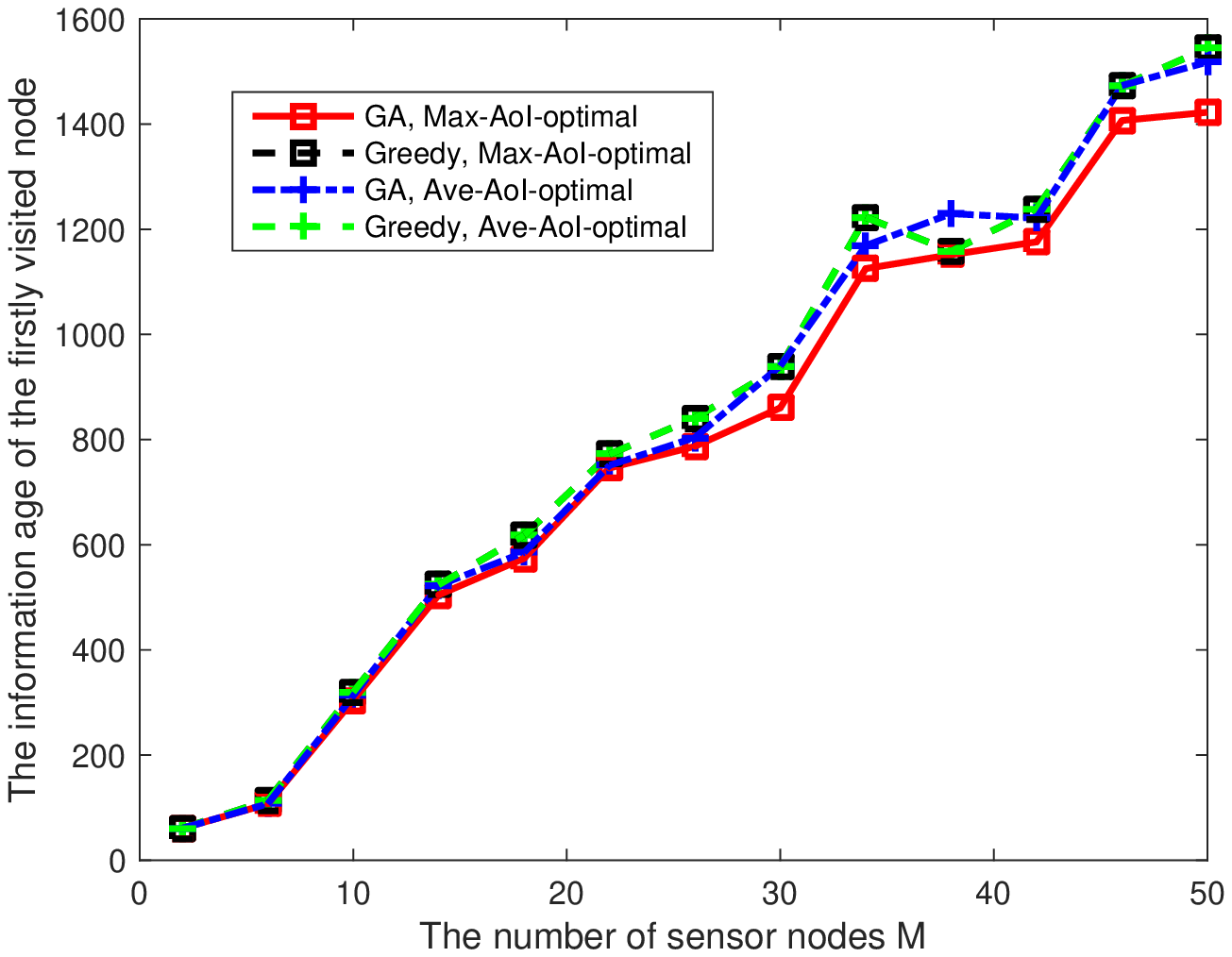}\caption{The minimum AoI $X_{1}^{*}$ versus the number of SNs $M$ in the
Max-AoI-optimal and Ave-AoI-optimal trajectories. \label{fig:age_X1_M}}
\end{figure}
\begin{figure}[t]
\centering \includegraphics[width=0.42\textwidth]{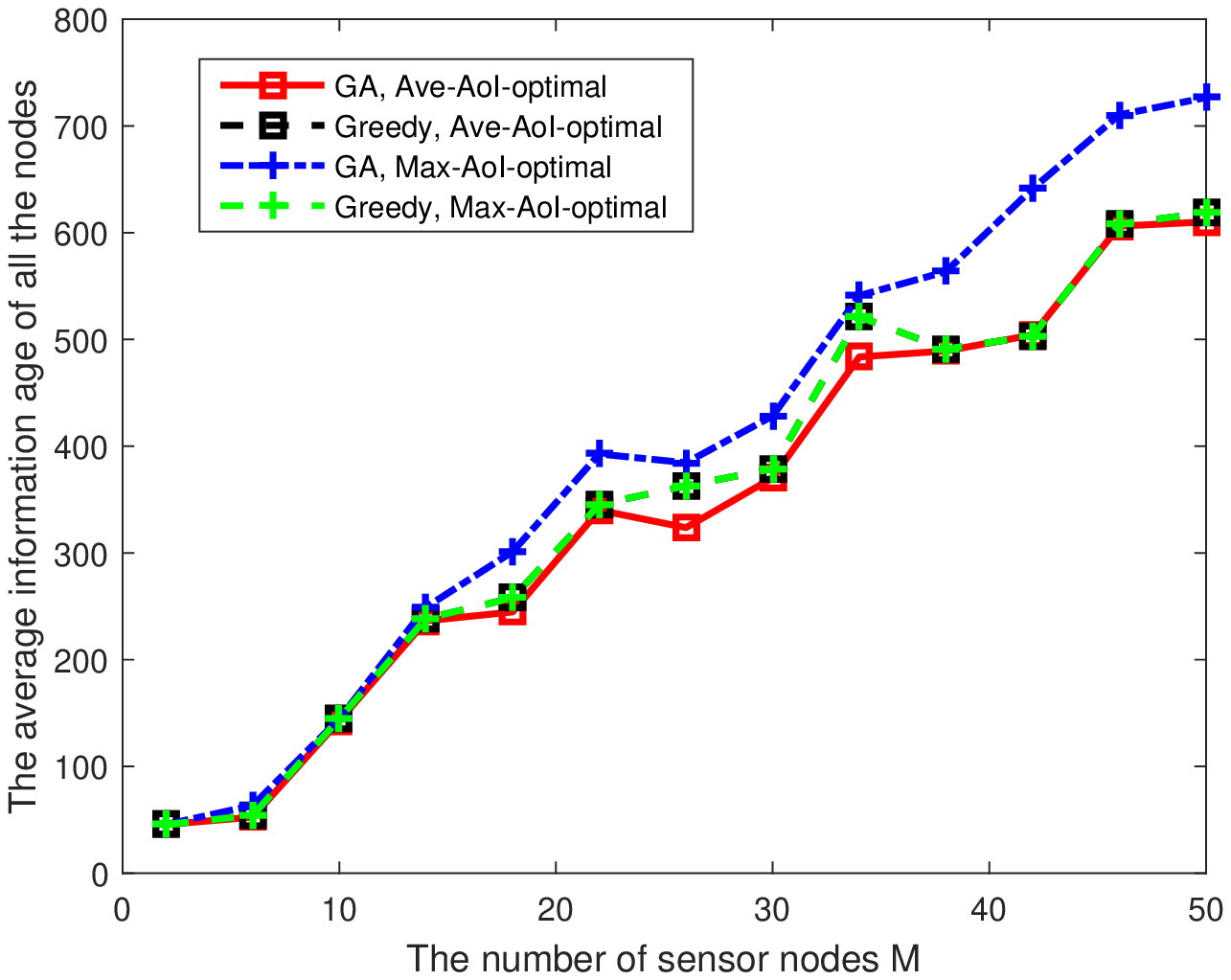}\caption{The minimum average AoI $\overline{X}_{1}^{*}$ versus the number
of SNs $M$ in the Max-AoI-optimal and Ave-AoI-optimal trajectories.\label{fig:average_age_X1_M}}
\end{figure}

\section{Simulation Results\label{sec:Simulation-Results}}

In this section, we present simulation results to demonstrate the
performance of the two trajectory planning algorithms. We consider
a UAV-enabled wireless sensor network that consists of one data center,
one UAV and $M$ SNs. The nodes are randomly located in a circular
area of radius $1000$m. The UAV is dispatched to collect data from
all the SNs exactly once along the flight trajectory. The flight height
and velocity are set as $50$m and $20$m/s, respectively. The sensed
data of each SN is uploaded to the UAV via a LOS link. The system
bandwidth is equal to $B=5$ MHz, and the channel power gain at the
reference distance $d_{0}=1$ m is set as $\beta=-60$ dB. The node's
transmission power and the noise power are set as $P_{i}=0.1$ watt
and $\sigma^{2}=-110$ dBm, respectively. The data sensing time at
each node is assumed to be very small and negligible. Suppose that
the UAV takes off at time $T_{0}=0$.

In simulations, the Max-AoI-optimal and Ave-AoI-optimal trajectories
are found using the three algorithms as follows: 1) the DP-based algorithm,
i.e., Algorithm \ref{alg:DP_AoI_optimal}; 2) the GA-based algorithm,
i.e., Algorithm \ref{alg:GA_AoI_optimal}; 3) the greedy algorithm
that recursively finds the nearest predecessor $v_{(i-1)}$ for the
node $v_{(i)}$ at each stage $i=M+1,M,\cdots,2$. More specifically,
the node closest to the data center $v_{0}=v_{(M+1)}$ is selected
and marked as node $v_{(M)}$. Similarly, among the unmarked nodes,
the node nearest to node $v_{(i)}$ is selected as node $v_{(i-1)}$.
This iterative procedure repeats until all the $M$ nodes constitute
a Hamiltonian path. The GA related parameters are set as $N_{c}=10^{3}$,
$N_{g}=10^{4}$, $\alpha=2$, $\gamma_{c}=0.8$ and $\gamma_{m}=0.01$. 

In Fig.$\,$\ref{fig:stage_Max_AoI} and Fig.$\,$\ref{fig:stage_Ave_AoI},
we present the AoI $X_{i}^{*}$ of node $v_{(i)}$ in the Max-AoI-optimal
and Ave-AoI-optimal trajectories, respectively. In this simulation,
we consider a data collecting network with $M=14$. For comparison,
we also plot the minimum average AoI $\overline{X}_{i}^{*}$ versus
the node index $i$ in Fig.$\,$\ref{fig:stage_Ave_AoI}. From these
figures, the information ages $X_{i}^{*}$ and $\overline{X}_{i}^{*}$
monotonically decreases with the node index $i$, in accordance with
the results in Lemma \ref{lem:flight_max_aoi} and Eq. (\ref{eq:average_aoi_inequality}).
Among the three algorithms, the DP-based algorithm performs the best
in terms of the AoI $X_{1}^{*}$ and $\overline{X}_{1}^{*}$, since
it can find the Max-AoI-optimal and Ave-AoI-optimal trajectories by
comparing all the candidate Hamiltonian paths. By intelligent search,
the GA-based algorithm can find near optimal trajectories on which
$X_{1}$ and $\overline{X}_{1}$ are very close to the minimum ones
$X_{1}^{*}$ and $\overline{X}_{1}^{*}$. In contrast, the greedy
algorithm achieves the largest AoIs $X_{1}$ and $\overline{X}_{1}$,
since it just finds the local optimum at each stage. 

The AoI $X_{i}^{*}$ of node $v_{(i)}$ ($i=2,\cdots,M$) found by
the DP algorithm may not be the smallest, as shown in Fig.$\,$\ref{fig:stage_Max_AoI}.
In this figure, the ages of information $X_{i}^{*}$ for $i=4,\cdots,12$
calculated by the DP algorithm are larger than that by the GA algorithm.
Similarly, $X_{i}^{*}$ for $i=10,\cdots,14$ are larger than that
obtained by the greedy algorithm. This is because that any part of
the globally optimal Hamiltonian path is not necessary to be locally
optimal. Different from the Max-AoI-optimal trajectory planning case,
the ages of information $X_{i}^{*}$ and $\overline{X}_{i}^{*}$ collected
from the nodes in the Ave-AoI-optimal trajectory found by the GA and
DP algorithms are very close, and are less than that obtained by the
greedy algorithm, as shown in Fig.$\,$\ref{fig:stage_Ave_AoI}. This
implies that the AoI metrics play the very important role in the flight
trajectory planning. 

In Fig.$\,$\ref{fig:age_X1_M} and Fig.$\,$\ref{fig:average_age_X1_M},
we plot the minimum AoI of the firstly visited node $X_{1}^{*}$ and
the minimum average AoI $\overline{X}_{1}^{*}$, respectively, when
the Max-AoI-optimal and Ave-AoI-optimal trajectories are found by
the GA and greedy algorithms. In Fig.$\,$\ref{fig:age_X1_M}, the
information age $X_{1}^{*}$ in the Max-AoI-optimal trajectory is
smaller than that in the Ave-AoI-optimal trajectory, when the GA algorithm
is applied. Similarly, the Ave-AoI-optimal trajectory achieves a much
smaller average AoI $\overline{X}_{1}^{*}$ than the Max-AoI-optimal
trajectory. Again, this points out the importance of the AoI metric
in the flight trajectory planning. When the greedy algorithm is applied,
the Max-AoI-optimal and Ave-AoI-optimal trajectories are exactly the
same, since the greedy algorithm always selects the the nearest neighbor
among the candidate nodes at each stage in either the Max-AoI-optimal
or Ave-AoI-optimal trajectory. It is also shown that the greedy algorithm
performs better in finding the Ave-AoI-optimal trajectory than finding
the Max-AoI-optimal trajectory, due to the increasing weight with
the stage index. 

\section{Conclusions\label{sec:Conclusions}}

This paper studied the Max-AoI-optimal and Ave-AoI-optimal trajectory
planning problems for UAV-enabled data collection in wireless sensor
networks. It was shown that the two age-optimal trajectories are exactly
two shortest Hamiltonian paths in a weighted complete graph. Then,
we developed DP and GA based algorithms to find Max-AoI-optimal and
Ave-AoI-optimal trajectories in a unified way. By simulations, we
showed that the proposed algorithms can find the age-optimal trajectories
efficiently, compared to the baseline greedy algorithm. Based on the
two AoI metrics, the UAV's trajectory design helps to keep the sensed
data fresh in wireless sensor networks. 

\bibliographystyle{IEEEtran}

\end{document}